\documentstyle[twocolumn,aps,prl,epsfig]{revtex}

\begin{document}
\draft

\twocolumn[\hsize\textwidth\columnwidth\hsize\csname@twocolumnfalse\endcsname%

\title{Triggered single photons from a quantum dot}

\author{Charles~Santori, Matthew~Pelton, Glenn~Solomon$^{*}$,
        Yseulte~Dale, and Yoshihisa~Yamamoto$^{\dagger}$}

\address{Quantum~Entanglement~Project, ICORP, JST,
         E.L.~Ginzton Laboratory, Stanford University,
         Stanford, California 94305}

\date{Submitted to \prl on October 9, 2000}

\maketitle

\begin{abstract}
We demonstrate a new method for generating triggered single photons.
After a laser pulse generates excitons inside of a single quantum dot,
electrostatic interactions between them and the resulting spectral
shifts allow a single emitted photon to be isolated. 
Autocorrelation measurements show a
reduction of the two-photon probability to 0.12 times the value for
Poisson light.  Strong anti-bunching persists when the emission is
saturated.  The emitted photons are also polarized.
\end{abstract}

\pacs{PACS numbers: 42.50.Dv, 78.66.-w, 73.23.-b}

\vskip2pc]

Photons from classical light sources, which usually consist of a
macroscopic number of emitters, follow Poisson statistics
or super-Poisson statistics~\cite{walls}. With a single quantum emitter,
however, one can hope to generate a regulated photon stream, containing
one and only one photon in a given time interval.  Such an ``anti-bunched''
source would be useful in the new field of quantum cryptography, where
security from eavesdropping depends on the ability to
produce no more than one photon at a time~\cite{qcrypt1,qcrypt2}.

Continuous streams of anti-bunched photons were first observed from single atoms
and ions in traps~\cite{atomion1,atomion2}.  More recently, experiments demonstrating
triggered single photons have used single molecules as the emitters, excited optically
either by laser pulses~\cite{pulsemolec,lounis} or through adiabatic following~\cite{brunelorrit}.

Solid-state sources have potential advantages.  Most importantly, they
may be conveniently integrated into larger structures, such as distributed-Bragg-reflector
(DBR) microcavities~\cite{dbr1,dbr2} to make monolithic devices.  In
addition, most do not suffer from the photo-bleaching effect that severely limits
the lifespan of many molecules.  The first experimental effort towards a solid-state
single-photon source was based on electrostatic repulsion of single carriers in a
semiconductor micropost p-i-n structure~\cite{jungsang}.  Milli-Kelvin temperatures
were required, however, and sufficient collection efficiency to measure the photon
autocorrelation function was not obtained.
More recently, continuous anti-bunched fluorescence has been seen from color centers in
a diamond crystal~\cite{diamond1,diamond2} and from CdSe quantum dots~\cite{cdse}.

Our method to generate triggered single photons involves pulsed optical excitation of
a single quantum dot and spectral filtering to remove all but the last
emitted photon.  Optically active quantum dots confine electrons and holes to small
regions so that their energy levels are quantized~\cite{qdgeneral}.  If several electrons or
holes are placed in the dot at the same time, they will, to a first approximation, occupy
single-particle states as allowed by the Pauli exclusion principle.  However, electrostatic
interactions between the particles cause perturbations in the eigenstates and energies.
For example, if two electron-hole pairs (excitons) are created (a ``biexcitonic'' state),
the first pair to recombine emits at a slightly lower energy than the second pair,
due to a net attractive interaction~\cite{biex1,biex2}.  We exploit this effect
to generate single photons not only through regulated absorption, as in the
single-molecule experiments, but also through this emission property, that the last
photon to be emitted after an excitation pulse has a unique wavelength, and therefore can
be spectrally separated from the others.

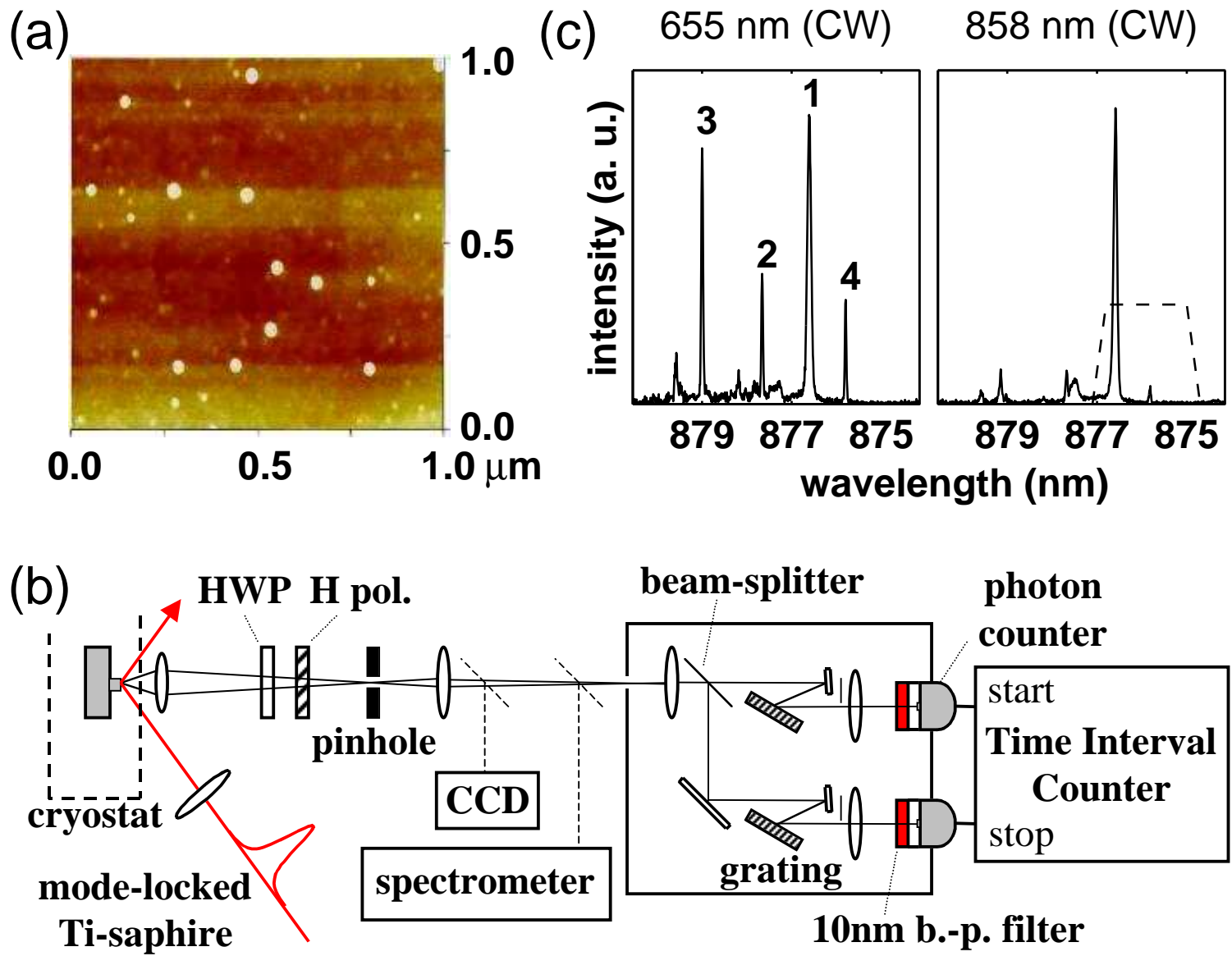
\begin{figure}
\caption{
(a)
Atomic force microscope image of uncovered InGaAs
self-assembled quantum dots, grown under identical conditions to those used in this
experiment.
(b)
The experimental setup, showing the laser-excited sample (left), collection
optics (middle-left), and Hanbury Brown-Twiss configuration (right).
(c)
Emission spectra from a quantum dot under above-band excitation (left)
and resonant excitation (right).  The dotted line indicates approximately the
portion of the spectrum that reaches the photon counters after filtering.}
\label{fig1}
\end{figure}

A sample was fabricated containing self-assembled InAs quantum dots
surrounded by a GaAs matrix~\cite{qdgeneral}. The dots were grown by molecular beam epitaxy
(Fig.~\ref{fig1}a) at a high temperature to allow alloying with the surrounding
GaAs, thereby shortening the emission wavelength.  They were
then capped by 75 nm of GaAs.  Mesas about 120~nm tall, 200~nm wide, and spaced
50 $\mu$m apart were fabricated by electron-beam lithography and dry etching.  The dots
are sparse enough (11 $\mu$m$^{-2}$) that the smallest mesas contain, on average,
fewer than one dot.

The experimental setup is shown in Fig.~\ref{fig1}b.  The sample was
cooled to 5K in a cryostat and placed close to the window.  A
mode-locked Ti-sapphire laser with 2.9 ps pulses and a 76 MHz
repetition rate was focused onto a mesa from a steep (53.5$^\circ$
from normal) angle, down to an 18 $\mu$m effective spot diameter.
Emission from the dot inside the mesa was collected with an NA=0.5
aspheric lens, and focused onto a pinhole that effectively
selected a 5 $\mu$m region of the sample for collection.  A
rotatable half-wave plate followed by a horizontal polarizer
selected a particular linear polarization. The light
was then sent to a CCD camera, a spectrometer, or a Hanbury
Brown-Twiss configuration for measuring the photon autocorrelation
function.  Two EG\&G ``SPCM'' photon counters were used for
detection, with efficiencies of 40\% at 877~nm,
and 0.2~mm-wide active areas.  A monochrometer-type
configuration defined a 2~nm-wide measurement bandwidth, with the
center wavelength determined by the detector position.  Additional
rejection of unwanted light (scattered pump light and stray room
light) was obtained with a 10~nm bandpass filter attached to each
detector.  The electronic pulses from the photon counters were
used as start ($t_1$) and stop ($t_2$) signals for a time interval
counter, which recorded these intervals $\tau=t_2-t_1$ as a
histogram.

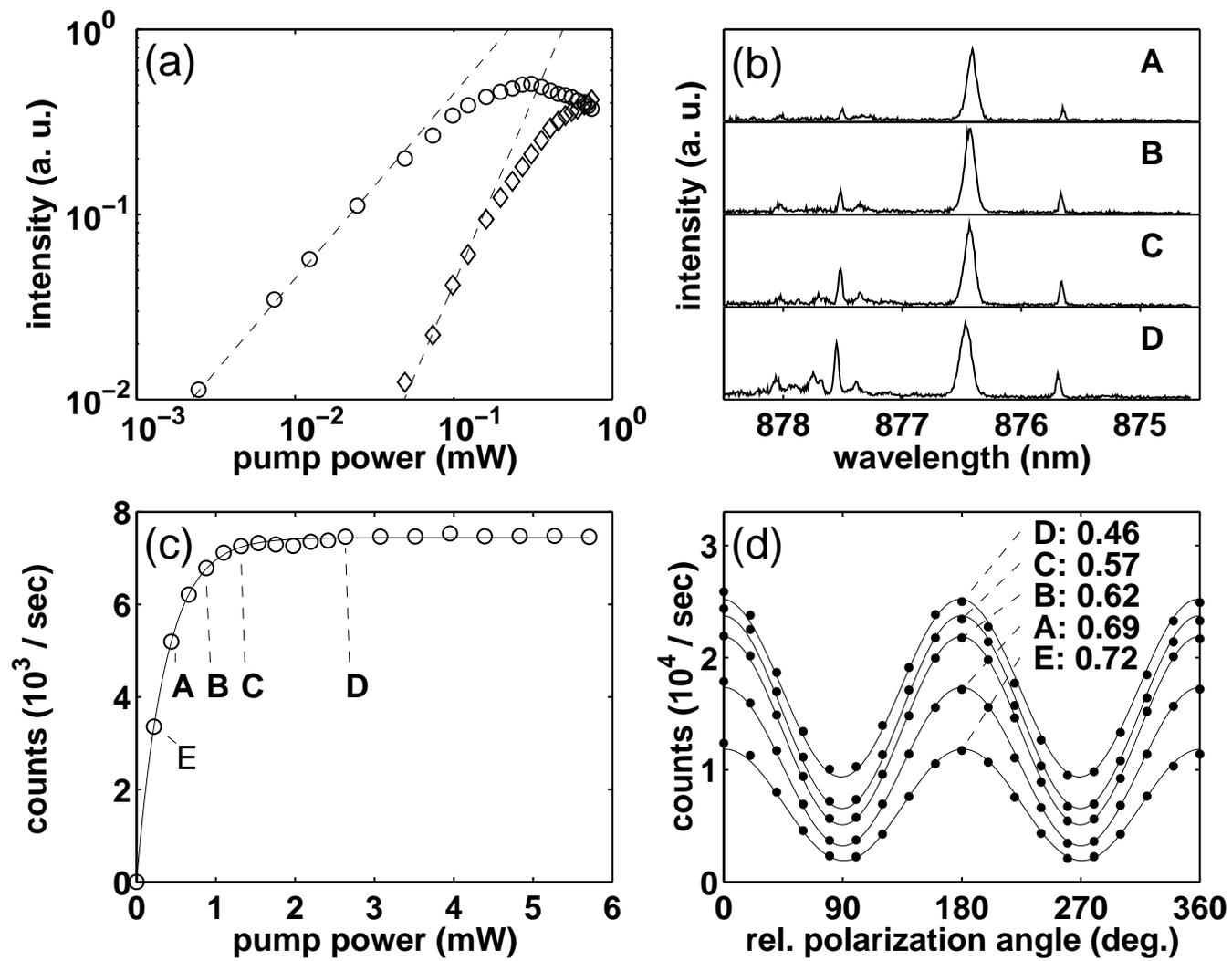
\begin{figure}
\caption{
(a)
A log-log plot of emission line intensity versus above-band (CW)
pump power, showing linear growth of peak~1 (circles) and
quadratic growth of peak~2 (diamonds).
(b)
Emission spectra,
(c) emission intensity (measured under the filtering condition
depicted in Fig.1b), and
(d) emission polarization dependence
of the dot under pulsed, resonant excitation with powers
E,A-D: 0.22~mW, 0.44~mW, 0.88~mW, 1.32~mW and 2.63~mW, respectively.
The count rates in (c) are further reduced by an additional
bandpass filter.  The solid line in (c) is a least-squares fit
of Eq.~\ref{satequ}, while the solid lines in (d)
fit a sinusoid plus an offset, resulting in the shown
visibilities (max.-min.)/(max.+min.).}
\label{fig2}
\end{figure}

Mesas containing single dots were identified by their optical emission spectra.
The mesa chosen for this experiment contains a dot whose main ground-state
emission wavelength is 876.4~nm.  With continuous-wave (CW) excitation above the GaAs
bandgap, the emission spectrum (Fig.~\ref{fig1}c, left) displays several lines,
as has been reported elsewhere~\cite{pistol}.  We believe that these lines all come from a
single dot because another mesa shows a nearly identical emission pattern (peak heights,
spacings and widths), except for an overall wavelength shift, suggesting that this pattern
is not random.  To avoid ionization of the dot or delayed capture of electrons and
holes, we tuned the laser wavelength to an absorption resonance at 857.5~nm, thus creating
excitons directly inside of the dot.  With resonant excitation, emission peaks~3 and~4
almost disappear (Fig.~\ref{fig1}c, right), and therefore we believe that they represent
emission from other charge states of the dot~\cite{chargedot}.  We identify peak~1 as
ground-state emission after the capture of a single exciton, and peak~2 as ``biexcitonic''
emission after the capture of two excitons.  This assignment is supported by the dependence of
the emission line intensities on pump power~(Fig.~\ref{fig2}a), showing linear growth
of peak~1 and quadratic growth of peak~2 in the weak pump limit.  A biexcitonic energy
shift of 1.7~meV is seen.

Under pulsed, resonant excitation, a clear saturation behavior is seen for
peak~1 (Fig.~\ref{fig2}b).  Although peak~2 and its surrounding peaks
(presumably multiple-excitonic emission) continue to grow as the pump
power is increased, peak~1 reaches a maximum value, since only the
last exciton to recombine emits at this particular wavelength.  This
is shown quantitatively in Fig.~\ref{fig2}c.  Here, a photon counter
was used to measure the emission rate versus pump power,
with the detection band tuned to accept peak~1 but reject peak~2
(see dashed line, Fig.~\ref{fig1}c).  A simple saturation
function for unregulated absorption that fits the data well is
\begin{equation} \label{satequ}
I = I_0 (1-e^{-P/P_{sat}}) \, ,
\end{equation}
where $I$ is the measured intensity for single-exciton emission,
$P$ is the pump power, and $I_0$ and $P_{sat}$ are fitting parameters
that characterize the total collection efficiency and the absorption
rate, respectively.

The emission from peak 1 was also linearly polarized.  Since the
degree of polarization of the emission depended strongly on the
pump polarization angle, we believe that the
effect is largely due to the selection rules for
photon absorption and emission~\cite{selection1,selection2}. The polarization
of a pump photon is transferred into the spin of an exciton, and
if no spin relaxation occurs, the spin is transfered back to the
emitted photon polarization.  The polarization is linear, as would
be expected for asymmetric dots under no magnetic
field~\cite{linearpol1,linearpol2}.  At the optimal pump polarization used in
this experiment, emission polarization with up to 72\% visibility
was observed at weak pump (Fig.~\ref{fig2}d).  The lack of perfect visibility
was perhaps due to spin relaxation, imperfect selection rules, or
effects of the post geometry.  The visibility was partially degraded when
the pump power was increased into the saturation regime.

We next examine the second-order coherence function, $g^{(2)}(\tau)$,
which contains information on photon emission statistics~\cite{walls}.
For a pulsed source, $g^{(2)}(\tau)$ becomes a series of peaks separated
by the laser repetition period, and the areas of these peaks give information
on photon number correlations between pulses separated by time $\tau$.  Of
special interest is the central peak at $\tau = 0$, which gives an upper
bound on the probability that two or more photons are emitted
from the same pulse:
\begin{equation}
\label{twophotbound}
2 P(n_j \geq 2) / \left<n\right>^2 \leq \frac{1}{T}
           \int^{\epsilon}_{-\epsilon}g^{(2)}(\tau) d\tau \, ,
\end{equation}
where $n_j$ is the number of photons in pulse $j$, $\epsilon$ is chosen to include
the entire central peak in the integration region, and $T$ is the pulse repetition period.
This result, along with $g^{(2)}(\tau)$, is independent of the collection and detection
efficiencies. For a ``classical'' (Poisson) source, the normalized central peak area
(right-hand side of Eq.~\ref{twophotbound}) is one.

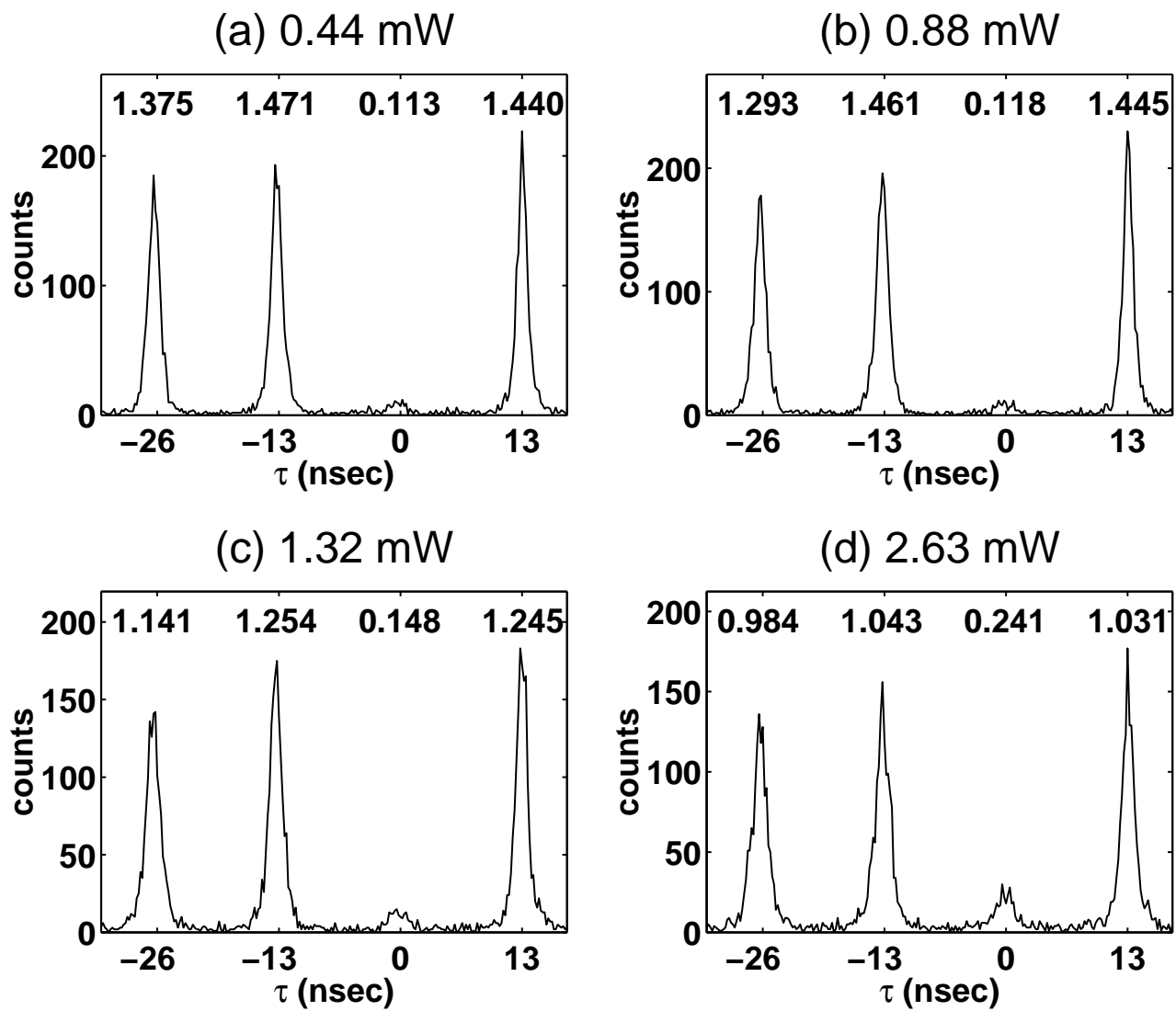
\begin{figure}
\caption{
Histograms of the time intervals $\tau=t_2-t_1$ between photons detected by the ``start''
and ''stop'' counters, for four different excitation powers: (a) 0.44~mW,
(b) 0.88~mW, (c) 1.32~mW and (d) 2.63~mW.  The numbers
printed above the peaks give the normalized autocorrelation peak areas,
calculated using a 5.6~ns-wide integration window.
The reduction of the $\tau=0$ peak demonstrates anti-bunching.}
\label{fig3}
\end{figure}

Histograms of the time interval $\tau=t_2-t_1$ taken at four different pump powers are
shown in Fig.~\ref{fig3}.  In the limit of low collection and detection efficiency ($\approx
0.0003$ combined in our case), these histograms, after correct normalization, approximate
the autocorrelation function.  The peaks are broader than the limit imposed by the photon
counter timing resolution (0.3~ns), and indicate a lifetime for the single-exciton state
of about 0.7~ns.  The $\tau=0$ peak shows a large reduction in area, indicating strong
anti-bunching.  The numbers printed above the peaks indicate the peak areas, properly
normalized by dividing the histogram areas by both singles rates, the laser repetition
period, and the measurement time.  For the numbers shown, the only background counts
subtracted were those due to the known dark count rates of the photon counters
(130~s$^{-1}$ \, and 180~s$^{-1}$), almost negligible compared to the singles rates,
19800~s$^{-1}$ \, and 14000~s$^{-1}$ \, for the two counters at 0.88~mW pump power.
When only counts within 2.8~ns of $\tau=0$ were included, a normalized $g^{(2)}(\tau=0)$
peak area of 0.12 was obtained at 0.88~mW. Subtracting the constant background floor
seen in the data gave an even lower value of 0.095.

The observed anti-bunching has two causes.  The first cause is a
suppression of the probability for the dot to absorb a second
photon after the first photon has been absorbed.  If one collects
emission from both the single-exciton and multi-exciton lines, the
$g^{(2)}(\tau=0)$ peak area is still reduced to about 0.32
due to limited absorption.  A possible explanation for
reduced absorption of the second photon is that electrostatic
interactions, similar to those responsible for the 1.7~meV
biexcitonic energy shift, move the absorption resonance to a lower
energy for the second photon \cite{twophoton}.
The second cause for the
observed anti-bunching is that, even if more that one exciton is
created, only the last exciton to recombine emits at our
collection wavelength.  Under these collection conditions, we see
a fair degree of anti-bunching at all pump wavelengths, even
above band, if the pump power is not too high.  The remaining
counts seen at $\tau=0$ under optimal pump and collection conditions
are most likely due to imperfect filtering to reject multi-excitonic
emission.

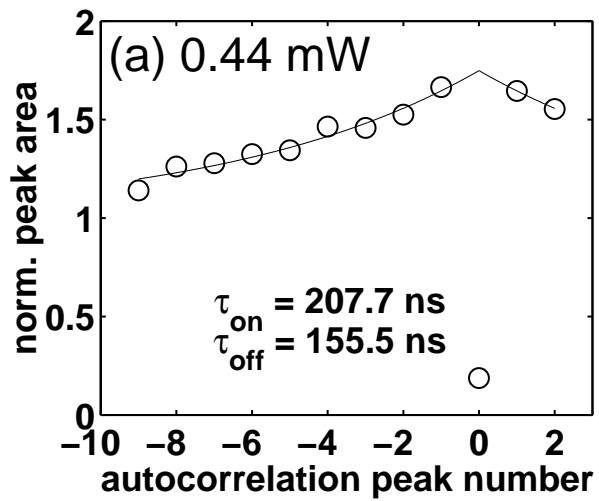
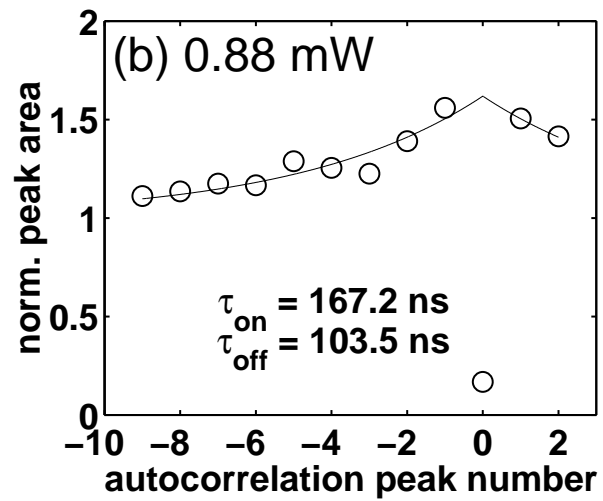
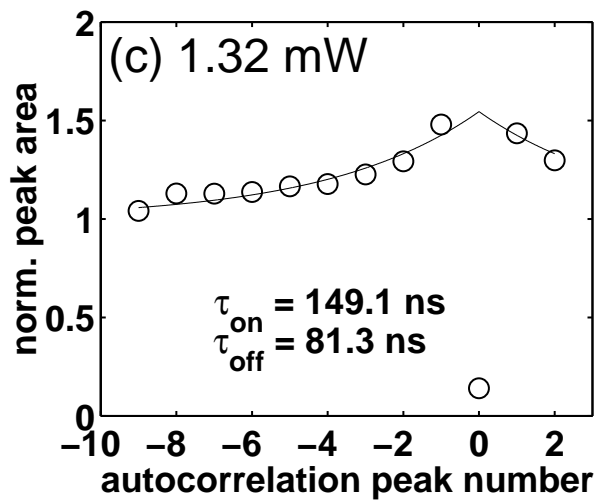
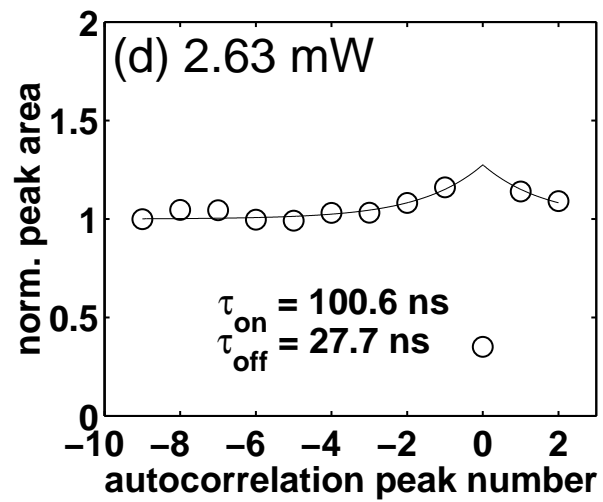
\begin{figure}
\caption{
Normalized autocorrelation peak areas (13~ns-wide integration window)
obtained from longer time-scale histograms, plotted against peak number, counted
from $\tau=0$, for four different excitation powers: (a) 0.44~mW,
(b) 0.88~mW, (c) 1.32~mW and (d) 2.63~mW.  The lines are
least-squares fits using Eq.~\ref{blinkmodl}, and the fitting
parameters obtained are shown.}
\label{fig4}
\end{figure}

While the central autocorrelation peak area is reduced, the
adjacent peaks have normalized areas larger than one.  This
indicates positive correlations between the detected photon numbers from
adjacent pulses.  This longer-term bunching behavior is better
seen in Fig.\ref{fig4}, which plots normalized autocorrelation
peak areas versus peak number over a longer time span.
The extra peak area above one is seen to decay
exponentially away from $\tau=0$. For larger pump
powers, the time scale and the magnitude of the effect decrease.  A
simple model to describe this behavior assumes that the dot
randomly ``blinks'' between two conditions, a fully functioning
condition and a ``dark'' (or wavelength-shifted) condition in which
photons are not observed, with time constants $\tau_{on}$ and $\tau_{off}$.
This model results in
\begin{equation} \label{blinkmodl} h_{m \neq 0} = 1 +
\frac{\tau_{off}}{\tau_{on}}e^{-(1/\tau_{off}+1/\tau_{on})|mT|} \,
,
\end{equation}
where $h_m$ is the $m$-th normalized autocorrelation peak area, and $T$ is the
laser repetition period.  Fitting this model to the data gives the values for $\tau_{on}$
and $\tau_{off}$ shown on the plots, which are on the order of 100~ns.
Long-term~($> 1 {\rm s}$) blinking behavior has already been reported in
strain-induced GaAs dots~\cite{gaasblink} and InP dots~\cite{inpblink}, and emission
wavelength fluctuations have been reported for InGaAs dots~\cite{inalasblink}.
These effects have been attributed to nearby defects~\cite{inpblink} and
trapped charges~\cite{inalasblink}.  The more rapid blinking behavior seen
here is unwanted and necessarily decreases the efficiency of the device, but
it should be contrasted with the bleaching behavior of single molecules.
The quantum dot described here has been studied for months and cooled
down to 5K about 30 times without ceasing to function or changing significantly.

The internal efficiency of single-photon generation is difficult to determine
because the collection efficiency of the first lens depends on uncertain factors
such as the position of the dot within the mesa and the quality of the surface.
An upper limit of 0.57 at 2.63~mW assumes that once the emission is saturated,
the only reductions are due to imperfect polarization and the $\approx$100~ns
blinking behavior described above. This would imply a collection efficiency
through the first lens of 0.006.  To improve the collection efficiency, a
realistic solution is to place the dot inside of a DBR microcavity to direct
most of the spontaneous emission into a single mode~\cite{dbr1,dbr2}.

In summary, we have demonstrated a new method for generating triggered single photons,
using a single quantum dot excited on resonance by laser pulses.  The method takes advantage of
Coulomb interactions between excitons and the resulting spectral shifts to isolate single
emitted photons.  We observed a ten-fold two-photon probability suppression and
strongly polarized emission, suggesting that a single quantum dot is a promising
candidate for a practical single-photon source, although some unwanted blinking
was also observed.  The main remaining challenge is to improve the collection efficiency,
which we expect can be accomplished by growing a microcavity around the dot.

We thank S. Somani for loaning critical equipment.
Financial assistance for C.~M.~S. was provided by the National Science
Foundation.  Financial assistance for C.~M.~S. and M.~P. was provided by Stanford
University.

\end{document}